# The Interplay Between Liquid-Liquid Phase Equilibria, Sequence, and $T_g$ in Copolymers

Makayla R. Branham-Ferrari[a], David S. Simmons[a*]

[a]Department of Chemical, Biological, and Materials Engineering, The University of South Florida, Tampa, Florida
[*]dssimmons@usf.edu

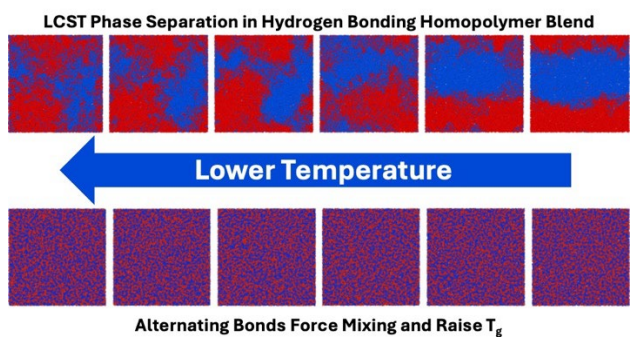

**Abstract**. Copolymerization is commonly employed to tune polymers' glass formation and improve properties such as ion conductivity and adhesion. Classically, mixing rules such as the Fox equation are employed to explain glass transition temperature ($T_g$) variations with copolymer composition. However, many copolymers deviate from these mixing rules in a manner that is monomer-sequence sensitive. We perform molecular dynamics simulations to probe the interplay between copolymer sequence, liquid-liquid phase equilibria, and $T_g$. We find that the direction and sequence-dependence of $T_g$ shift are predicted by the liquid-liquid phase behavior of the comonomers. Systems tending towards Upper Critical Solution Temperature behavior negative $T_g$ deviations, while systems tending towards Lower Critical Solution Temperature behavior exhibit positive $T_g$ deviations. In both cases, this effect is strengthened with increasing alternation – a consequence of bond-induced forced mixing. These results inform strategies for rationally varying copolymer $T_g$, at fixed composition, via design of polymer chain sequence.

## Introduction

The glass transition exerts enormous control over polymers' performance, including thermal stability ranges, transport capabilities, and mechanical properties.[1] As one of many examples, promising bio-derived polymers such as PLA commonly face limitations due to insufficiently high glass transition temperature $T_g$ and resulting limitations on the thermal range accessible to applications. When an underlying polymer's chemical and economic attributes are favorable, a major approach to overcoming this type of challenge is augmentation via blending,[2,3] introduction of additives,[4–24] or copolymerization.[25–34] However, these approaches face at least two crosscutting challenges. First, they involve changing the chemical composition of the polymer, such that desirable chemical properties may be attenuated even as $T_g$ and associated physical properties are improved. Second, it is generally far more challenging to *enhance* the $T_g$ of a composite or copolymer above the value of its components than it is to suppress $T_g$ – a consequence of both foundational $T_g$ mixing rules and of the thermodynamic tendency of intercomponent interactions to be weaker than self-interactions.

Indeed, the $T_g$ of a statistical copolymer most commonly lies between or modestly below the two component $T_g$s, as modeled by the Fox equation. This equation uses two components' homopolymer $T_g$s and respective weight fractions to predict a mass-fraction-weighted inverse-averaged mixture $T_g$ such that

$$\frac{1}{T_g} = \frac{w_1}{T_{g,1}} + \frac{w_2}{T_{g,2}} \quad (1)$$

where $w_i$ is defined as the weight fraction of monomer (or component) i and $T_{g,i}$ is the $T_g$ of the component *i*'s homopolymer $T_g$. In addition to the limited $T_g$ range of accessible $T_g$'s suggested by this model, random copolymers and blends often inherit undesirable properties from either component that can be difficult to decouple via composition alone.

Fortunately, experiments have observed a subset of systems in which $T_g$ can deviate significantly from the weighted average predicted by the Fox equation.[35–47] Indeed, positive deviations are sufficiently common[37,45,46] that empirical models such as the Kwei equation[45] or the Gordon-Taylor equation,[48] which include fitted interaction parameters, are commonly employed to describe these systems. However, these models are not predictive or physically explanatory – they involve fitting the blend or copolymer system to an empirical model. On the one hand, these deviations from Fox behavior present an enormous opportunity to access composites with properties more extreme than either component. On the other, they pose a fundamental question – how can we understand these deviations sufficiently well to enable their rational design?

Evidence suggests that the strength of cross-interactions between the species – closely related to the energy of mixing of the underlying homopolymers of the monomers comprising the copolymer – plays a central role in mediating the direction and magnitude of deviations from the Fox equation.[37] Indeed, it has long been recognized that systems involving hydrogen bonding between species have a greater tendency towards positive deviations from Fox mixing.[45] At a simplified level, these cross-interactions can be parameterized via the exchange energy, which reports the energy change associated with exchanging a pair of molecules of the two species between their pure-state melts. This quantity reflects the difference between self-interactions and cross-interactions, in which negative exchange energy represents cross-interactions that are stronger than self to self-interactions, and vice versa. Since short-range cohesive energy plays a leading order role in determining $T_g$,[49] it is reasonable to expect that favorable (negative) exchange energies will tend to yield enhancement in $T_g$ on mixing and vice versa. Indeed, many copolymers exhibiting deviations from Fox appear to accord with this expectation.[37,39,45,46,50]

As a particularly relevant example, previous work involving poly(vinylidene chloride-co-methyl acrylate) report deviations from Fox that are interesting in two ways.[37,51,52] This system consists of monomers with nearly identical homopolymer $T_g$s. With this system, researchers spanned composition fractions of random copolymers from pure P(VDC) to pure P(MA) and measured $T_g$. Rather than following the weighted average between the homopolymer $T_g$s that is predicted by Fox, the $T_g$ of the random copolymer reached a maximum at 50:50 VDC:MA composition fraction.[37] At equal VDC:MA composition, the $T_g$ of the random copolymer was nearly 40K above that of either homopolymer. Even more interestingly, the equal VDC:MA composition alternating copolymer has a $T_g$ that is 15K higher than the random copolymer.[37] This magnitude of $T_g$ enhancement, both from Fox and between sequences, provokes questions regarding the relationship between copolymer sequence and $T_g$. Conversely, $T_g$ suppression was seen in tapered block polymers comprised of poly(styrene-*b-oligo*-oxyethylene methacrylate) [P(S-OEM)] and employed by Epps et al. to increase transport properties for ion conduction.[53,54] Beyond mere Tg suppression, this study found that tapers vs reverse tapers led to considerably different dynamics, even at fixed overall composition. These results indicate that Fox deviations can be tuned via sequence to predictably achieve copolymer $T_g$s beyond that of the copolymer components.

Recent computational work by our group probed the molecular mechanisms of the latter case, in which copolymerization results in $T_g$ suppression relative to the Fox equation.[50] Drayer et al. reported on pronounced negative deviations from Fox in bead spring copolymers with a positive monomer-monomer exchange energy.[50] These simulations also found that these $T_g$ deviations are strongly mediated by sequence effects, with near-alternating polymers exhibiting much larger $T_g$ suppressions than blockier chains. At a molecular scale, this study indicated that covalent interconnectivity forced energetically unfavorable contacts between dissimilar monomers, reducing cohesive energy and $T_g$. More alternating sequences drive a stronger forced mixing effect of this kind, and thus lead to stronger $T_g$ suppressions.

That prior simulation work reported two regimes of sequence effects on $T_g$.[50] In systems with large blocks, sequence modifies domain size and thus modulates interfacially-induced gradients in relaxation time and $T_g$ that are characteristic of near-interface glassy dynamics more generally.[55–60] In systems with small block size (i.e. near-alternating systems), sequence modulates local segmental packing or mixing and thereby modulates the system's cohesive energy and $T_g$. This two-mechanism picture was found specifically in systems of weakly-interacting comonomers (i.e. those with unfavorable exchange energies) that are immiscible in the homopolymer-blend limit. This raises a series of questions regarding the technologically essential alternate case in which copolymer $T_g$ is enhanced relative to the Fox equation, opening the door to polymers with expanded thermal performance windows.

- How does sequence modulate $T_g$ in systems in which the segmental exchange energy is favorable? (i.e. in which their cross interaction is preferentially favorable)
- The mechanism described above effectively implicates A-B bonds in compelling forced mixing of otherwise immiscible monomers. How, then does this mechanism change in systems for which the underlying monomers are already well-mixed in the absence of A-B bonds, as might sometimes be expected in the presence of favorable cross-interactions?
- More broadly, this force-mixing scenario suggests a potential interplay between monomer liquid-liquid mixing behavior (e.g. liquid-liquid phase boundaries) and sequence effects on $T_g$? What is the nature of this interplay?

To answer these questions, here we employ molecular dynamics simulations to probe sequence effects on $T_g$ in copolymer systems in which

(i) the underlying homopolymers are immiscible due to an unfavorable exchange energy;
(ii) the underlying homopolymers are miscible due to a favorable exchange energy; and
(iii) the monomers possess a favorable exchange energy but exhibit phase separation due to an unfavorable entropy of mixing.

These three systems correspond to (i) a pair of monomers that tend to exhibit an upper critical solution temperature (UCST), (ii) a pair of monomers that are fully miscible at any block molecular weight, and (iii) a pair of monomers that tend to exhibit a lower critical solution temperature (LCST).

Results of these simulations indicate that liquid-liquid phase equilibria play a central role in determining the sequence-dependence of $T_g$ in copolymers. As seen in the work of Drayer et al.[61], homopolymer systems with negative exchange energy yield copolymers that exhibit a suppressed $T_g$ relative to the Fox rule, with this suppression growing as block lengths drop towards the alternating limit. Intuition suggests that opposite exchange energies should lead to opposite deviations from the Fox prediction. Indeed, in systems exhibiting positive exchange energy due to favorable van der Waals cross interactions, copolymer $T_g$ is enhanced relative to the Fox rule. However, $T_g$ enhancement *weakens* in these systems' alternating limit, rather than *strengthening* the underlying $T_g$ shift as in the negative exchange energy case. We find that this behavior results from the absence of a forced-mixing effect in near-alternating chains, because the underlying homopolymers are unconditionally miscible. In contrast, we find that when positive exchange energies emerge from strong directional interactions (e.g. hydrogen bonding) as is more common for experimental polymers, $T_g$ is both enhanced on copolymerization and exhibits further forced-mixing enhancement in the alternating limit. We find that this emanates from LCST-driven demixing of the underlying monomers in the long-block limit, as typically occurs in the presence of high molecular weight polymers with strong directional interactions.[62,63] These results point to a complex interplay of entropy and enthalpy of mixing, phase boundaries, sequence effects, and $T_g$ in copolymers.

Table 1. Interaction parameters for forcefields employed in simulated systems

|     | Bead I | Bead J | ε | σ | $r_{cut}$ |
|-----|--------|--------|---|---|-----------|
| B-B | 1 | 1 | 1 | 1 | $2.0\sigma_{BB}$ |
| B-C | 1 | 2 | 0.87 | 1 | $2.0\sigma_{BC}$ |
| B-A | 1 | 3 | 1 | ½ ($\sigma_{BC}$ + $\sigma_{AD}$) | $2^{(1/6)}\sigma_{BA}$ |
| B-D | 1 | 4 | 1 | ½ ($\sigma_{BC}$ + $\sigma_{AD}$) | $2^{(1/6)}\sigma_{BD}$ |
| C-C | 2 | 2 | 1 | 1 | $2.0\sigma_{CC}$ |
| C-A | 2 | 3 | 1 | ½ ($\sigma_{BC}$ + $\sigma_{AD}$) | $2^{(1/6)}\sigma_{CA}$ |
| C-D | 2 | 4 | 1 | ½ ($\sigma_{BC}$ + $\sigma_{AD}$) | $2^{(1/6)}\sigma_{CD}$ |
| A-A | 3 | 3 | 0 | $2.3\sigma_{AD}$ | $2^{(1/6)}\sigma_{AA}$ |
| A-D | 3 | 4 | $\varepsilon_{AD}$ | .3 | $2\sigma_{AD}$ |
| D-D | 4 | 4 | 0 | $2.3\sigma_{AD}$ | $2^{(1/6)}\sigma_{AA}$ |

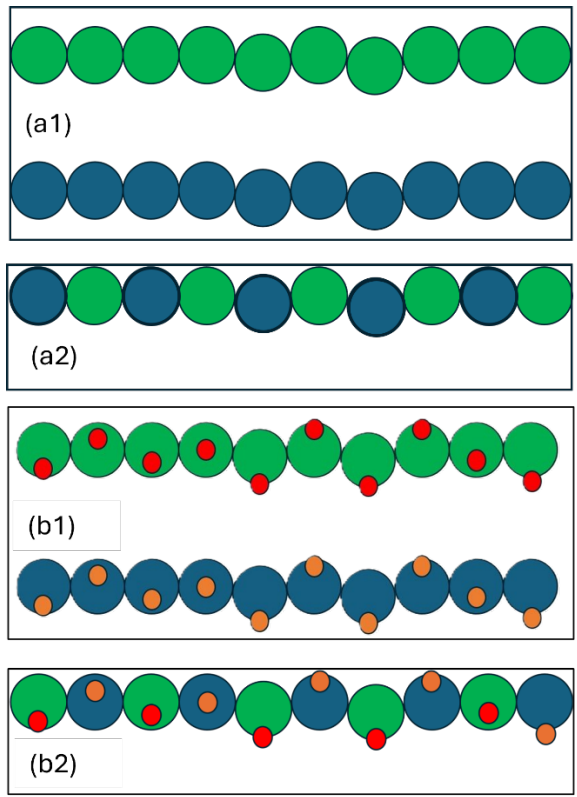

Figure 1. Schematic of hydrogen bonding bead spring model in the literature, showing a blend of homopolymers (a) and an alternating copolymer (b). The smaller embedded beads model hydrogen bonds (between red and orange beads) while the main backbone beads (green and blue) model dispersion interactions.

## Methods

We perform molecular dynamics simulations using distinct models for systems including only van der Waals interactions and those incorporating hydrogen bonding. In both cases, we perform simulations via the Large-Scale Atomic/Molecular Massively Parallel Simulator (LAMMPS) software package [64,65–69, 70], with GPU acceleration[71] and using PACKMOL[72] to generate initial conditions. All simulations employ the Nose-Hoover barostat/thermostat as implemented in LAMMPS to control temperature and pressure, with pressure P = 0.

For van der Waals interactions, we simulate bead-spring polymers interacting via an attractive variant Kremer-Grest bead-spring model. In this model, the underlying tendency of the monomers towards miscibility is controlled the interaction parameters for a 12-6 Lennard-Jones non-bonded potential, with cutoff distance $r_{cut}$ = 2.5 and size parameter σ = 1. Bead interaction strength parameters are set to $\varepsilon_{AA}$ = $\varepsilon_{BB}$ = 1 for self-interactions, and the cross-interaction parameter $\varepsilon_{AB}$ is varied to modulate the exchange energy and miscibility of polymer segments of types A and B. Bonded interactions employ the finitely extensible nonlinear elastic (FENE) potential with parameters K=30, $R_0$=1.5, ε=1, and σ=1. This potential is sufficient to implement the dispersion-type forces needed to realize models (i) and (ii) described in the introduction: (i) the underlying homopolymers are immiscible due to an unfavorable dispersive exchange energy; (ii) the underlying homopolymers are miscible due to a favorable dispersive exchange energy.

To model LCST behavior in a bead spring model, we must introduce additional interactions to reflect the underlying physics of polymer-polymer LCST behavior. Classic LCST behavior stems from a tradeoff between favorable energetics between monomer pairs and an entropic penalty of mixing.[73–75] This behavior is common in systems that include strong, directional interactions such as hydrogen bonding.[63] Formation of hydrogen bonds is energetically favorable but reduces the orientational degrees of freedom of hydrogen bonding groups and thus generates an unfavorable contribution to entropy.

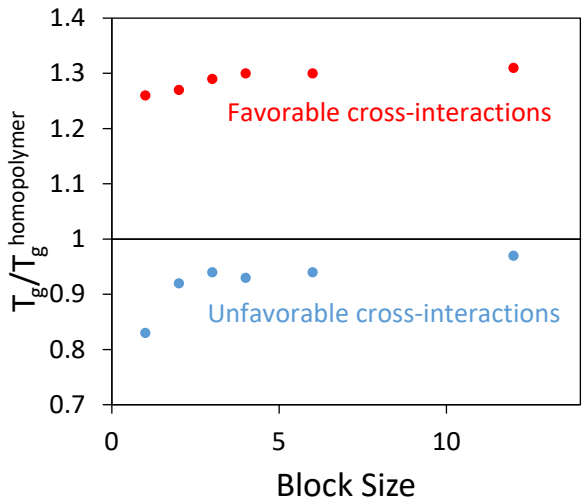

*Figure 2. $T_g$ vs blocksize for unfavorable (blue) and favorable (red) systems. Because the $T_g$s of both underlying homopolymers are equal, the horizontal line where $T_g/T_g^{homopolymer} = 1$ corresponds to the Fox prediction for all of these blends.*

Modeling this effect requires introduction of a forcefield with hydrogen-bond-like directional interactions. As adapted from Jarayaman et al., a simple model for hydrogen bonding can be created through the addition of "hydrogen bond beads" embedded into backbone beads (Figure 1).[76–79] The beads representing hydrogen bonds (HB) are a fraction of the size of the backbone beads (BB) with diameters of $0.3\sigma$ and $1.0\sigma$, respectively.[77,78] The HB beads are fixed to the BB beads via a short bond to place the HB bead on the surface of the BB bead. The embedded HB beads add effective directionality to the model via an angular potential on the short bond between HB and BB beads.[77,78] The bond angle is set to be very stiff relative to the backbone and restricts the HB bead to a 90° angle relative to the backbone.

Our model includes an array of non-bonded and bonded potentials (FENE) with varying strengths, lengths, and interaction groups, as shown in Table 1. Chains consist of backbone beads of types A and B as well as hydrogen bonding acceptor and donor beads, C and D, respectively, where acceptors are bound to type A backbones and vice versa. The FENE potentials for backbone bead to backbone bead are identical regardless of backbone bead type. The bonds between backbone beads and hydrogen bonding beads are much shorter than those between backbone beads. Non-bonded interactions use the LJ potential with an array of interaction parameters for different interactions. In the case of backbone interactions, A-A, B-B, or A-B, the potential includes neutral interactions with $\varepsilon_{AA} = \varepsilon_{BB} = 1.0$. Acceptor-acceptor or donor-donor interactions between hydrogen bonding beads (C-C and D-D) are set as neutral. Backbone beads and hydrogen bond beads do not undergo non-bonded interaction as $\varepsilon_{BB–HB} = 0$. The strength of the hydrogen bonds is set by the $\varepsilon_{HB}$ parameter controlling the interactions between acceptors and donors (C-D).

For each model, simulations are performed for a range of block sizes with fixed molecular weight and composition (50:50). We also perform simulations of each homopolymer to normalize the resulting $T_g$.

We employ a thermal annealing protocol defined by the Predictive Stepwise Quench Algorithm reported in prior work.[80] Within this approach, each system is subject to isothermal annealing periods of at least ten times the segmental relaxation time over a broad of temperature, to a target maximum relaxation timescale $\tau_{max}$ (corresponding to the lowest temperature simulated for a given system). We employ $\tau_{max} = 10^4 \tau_{LJ}$ (Lennard Jones time units, corresponding approximately to ps) for systems with dispersion interactions only, and $10^3 \tau_{LJ}$ for hydrogen bonding systems, which are more computational expensive to simulate.

### Simulation Analysis

We quantify segmental relaxation via the self-part of the intermediate scattering function $F_s(q,t)$. Here $q$ is the wavenumber, set to 7.07 as in a large body of prior work, comparable to the first structure factor peak. The relaxation time $\tau$ is defined as the time at which the $F_s(q,t)$ curve decays to 0.2, and we smooth and interpolate this time via a stretched exponential fit to the long-time portion of the relaxation function. We then obtain a computational timescale glass transition temperature as follows. We fit $\tau(T)$ to the cooperative model of Schmidtke et al.[81]. We then define a computational-timescale $T_g$ using a convention that $\tau(T=T_g) = \tau_{max}$ for that system. Each system is equilibrated via the predictive stepwise quenching (PreSQ) method[80] and analysis is performed using the Amorphous Molecular Dynamics Analysis Toolkit (AMDAT).[82,83]

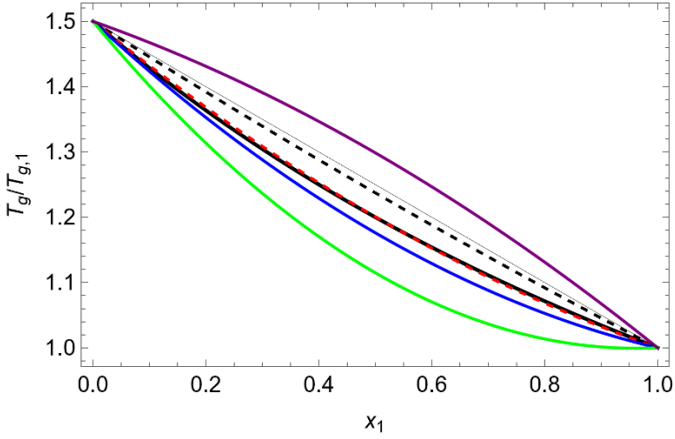

*Figure 3. Predictions of the Fox mixing rule (equation (1)) and a minimalist Berthelot-based cohesive energy $T_g$ mixing rule (equation (5)) for a copolymer or blend in which $T_{g,2} = 1.5 T_{g,1}$. The solid black curve is the Fox rule; the dashed black curve is equation (5) with $\kappa = 1$. The solid green, solid blue, dashed red, and solid purple curves are equation (5) with $\kappa = 0.8, 0.9, 0.94,$ and $1.1$, respectively. As shown by comparison of the dashed red and solid black curves, equation (5) with $\kappa = 0.94$ closely mimics the Fox rule with this $T_g$ ratio. The light black line is a linear mixing rule.*

## Results and Discussion

### Fox Mixing and deviations in model systems with vdW interactions only

Because these simulations are performed at a 50:50 composition of two monomers that comprise homopolymers with equal $T_g$ (i.e. $T_{g1}=T_{g2}$), the Fox rule predicts that dynamics and $T_g$ should be unaltered at any composition of this copolymer pair. $T_g/T_g^{homopolymer}$ then directly reports on deviations from the Fox rule. As shown in Figure 2, the Fox prediction dramatically breaks down in the presence of preferentially favorable or unfavorable cross-interactions between the two polymers. Consistent with our prior work, unfavorable cross-interactions (a positive exchange energy) lead to a suppression in $T_g$ relative to Fox.

Furthermore, we find that the direction of alterations in $T_g$ inverts when we invert the sign of the exchange energy by introducing preferentially favorable cross-interactions, as shown by the red datapoints in Figure 2. In this case, the copolymer exhibits an increase, rather than a decrease, in polymer $T_g$ relative to the homopolymer baseline.

If exchange energies have a leading-order influence the direction and magnitude of deviations from Fox mixing rules, this raises the question of why the Fox mixing rule should *ever* work reasonably well. Put another way, why is it the case that a large number of copolymers evidently exhibit an approximately neutral interaction for the purpose of glass formation?

To answer this, we sketch an extremely minimalist model for the impact of interaction strength on copolymer $T_g$. It is well established that, all else being equal, the strength of short-range cohesive interactions is central to determining $T_g$, with stronger cohesion leading to higher $T_g$. In systems dominated by dispersion forces, all cohesive energy is short ranged, such that we might very loosely sketch $T_g \sim E_c$, where $E_c$ is the cohesive energy. Within a Flory-Huggins like perspective, the mixing energy can be written as,

$$E_c \propto x_1^2 \varepsilon_{11} + x_2^2 \varepsilon_{22} + 2 x_1 x_2 \varepsilon_{12}, \qquad (2)$$

where $x_k$ is the lattice site fraction (approximately volume or weight fraction) of species $k$ and $\varepsilon_{jk}$ is the interaction energy parameter between sites of species $j$ and $k$. For systems involving only dispersion interactions, the cross-interaction parameter $\varepsilon_{12}$ is usually well-described by a Berthelot mixing rule,

$$\varepsilon_{12} = \kappa \sqrt{\varepsilon_{11} \varepsilon_{22}}, \qquad (3)$$

with Berthelot parameter $\kappa \cong 1$.

$$E_c \propto x_1^2 \varepsilon_{11} + x_2^2 \varepsilon_{22} + 2 x_1 x_2 \kappa \sqrt{\varepsilon_{11} \varepsilon_{22}}. \qquad (4)$$

If we then, as an extremely course approximation, assume $T_g \sim E_c$, it follows that

$$T_g \sim \left( x_1 \sqrt{T_{g,1}} + x_2 \sqrt{T_{g,2}} \right)^2 + 2(\kappa - 1) x_1 x_2 \sqrt{T_{g,1} T_{g,2}}. \qquad (5)$$

Notably, values of $\kappa$ are commonly close to but less than 1.0 for systems with vdW interactions only, with typical values as low as 0.8.[84] In this range of $\kappa$, equation (5) can closely mimic the Fox equation (equation (1)). For example, Figure 3 illustrates that equation (5) with $\kappa = 0.94$ closely mimics the Fox equation when $T_{g,2}/T_{g,1} = 1.5$.

To be clear, we do not suggest that $T_g$ is literally controlled purely by cohesive energy – this would be a dramatic oversimplification. However, this analysis suggests that simple cohesive energy effects can explain to leading order why the Fox equation frequently works reasonably well as an empirical matter, and why it frequently fails.

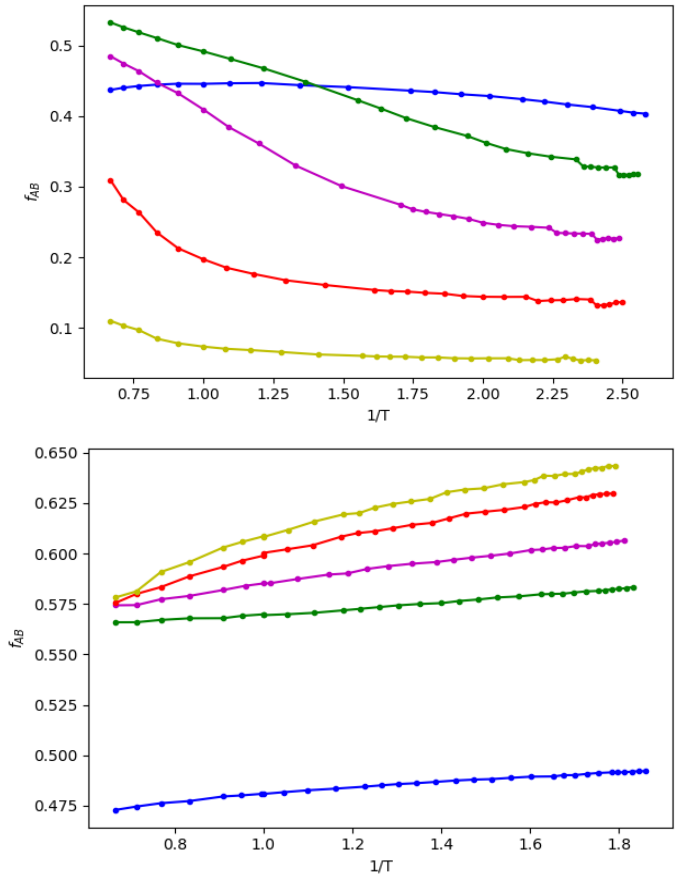

*Figure 4 Top: the fraction $f_{AB}$ of nonbonded opposite type neighbors (defined in the text) vs inverse temperature for systems with unfavorable dispersion interactions. Bottom: $f_{AB}$ vis inverse temperature for the systems with favorable vdW interactions. Systems correspond, from bottom to top in the upper figure and top to bottom in the lower figure, to systems with repeating block size 12 (yellow), 6 (red), 3 (magenta), 2 (green) and 1 (blue).*

Put simply, most vdW polymers mildly dislike each other energetically; inserting this fact into a simple cohesive energy mixing rule for $T_g$ yields a mild upward concavity to the $T_g$ mixing curve that is comparable to the predictions of Fox. Moreover, this model predicts that favorable cross-interactions at a van der Waals level should lead to enhancement and downward concavity of the $T_g$ mixing curve. This is consistent with the $T_g$ enhancement we reported with favorable vdW mixing energies in Figure 2.

## Sequence effects from Fox Mixing in model systems with vdW interactions only

While this mixing-energy scenario seems to accord with both the baseline Fox mixing rule and with the general direction of $T_g$ shifts with interaction energy, our simulations point to a surprising difference in the sequence-dependence of $T_g$ in the cases of unfavorable vs favorable exchange energy. In the case of unfavorable exchange energy, the sequence-driven deviation from Fox is maximized in the shortest block lengths, in the alternating limit. However, in the case of favorable exchange energy, the sequence-driven deviation from Fox is maximized in the long block length limit and minimized in the alternating limit (Figure 2).

Why does this occur? In our prior work, which focused on the unfavorable exchange energy case, we observed two mechanisms for sequence dependence. For large blocks, opposite type beads effectively separate to minimize energetic penalties and form interfaces between regions. For small blocks, the increase in A-B bonds force the system to mix and incur energetic penalties that accelerate dynamics and suppress $T_g$. In the case of favorable dispersive exchange energies, the former mechanism is always absent because these systems are unconditionally miscible and thus possess no well-defined internal domains or interfaces to nucleate such an interfacial effect. Therefore, the trend observed for favorable interactions in our simulations suggests that the forced segmental mixing mechanism is either absent or somehow operates in the reverse direction in this case.

To understand this, we compute the fraction $f_{AB}$ of nearest neighbors to a reference bead that are nonbonded and the opposite type,

$$f_{AB} = \langle m_{AB} \rangle - \langle m_{AB}^{bonded} \rangle / \langle m_{AB} \rangle + \langle m_{AA} \rangle - \langle m^{bonded} \rangle \quad (6)$$

where for a central A bead, $\langle m_{AB} \rangle$ is the average number of B neighbors, $\langle m_{AA} \rangle$ is the average number of A neighbors, $\langle m_{AB,bonded} \rangle$ is the average number of B neighbors that are connected to the central A bead via a bond, and $\langle m_{bonded} \rangle$ is the total average number of bonded neighbors. This metric gives insight into the local packing structure of the system, reporting on the extent to which a typical A bead contacts opposite-type rather than same-type neighbors. Because the structure and $T_g$ of A and B homopolymers and monomers are symmetric, and because we focus on 50:50 composition systems, $f_{AB} = f_{BA}$ for this system by construction and we thus focus on $f_{AB}$ for simplicity.

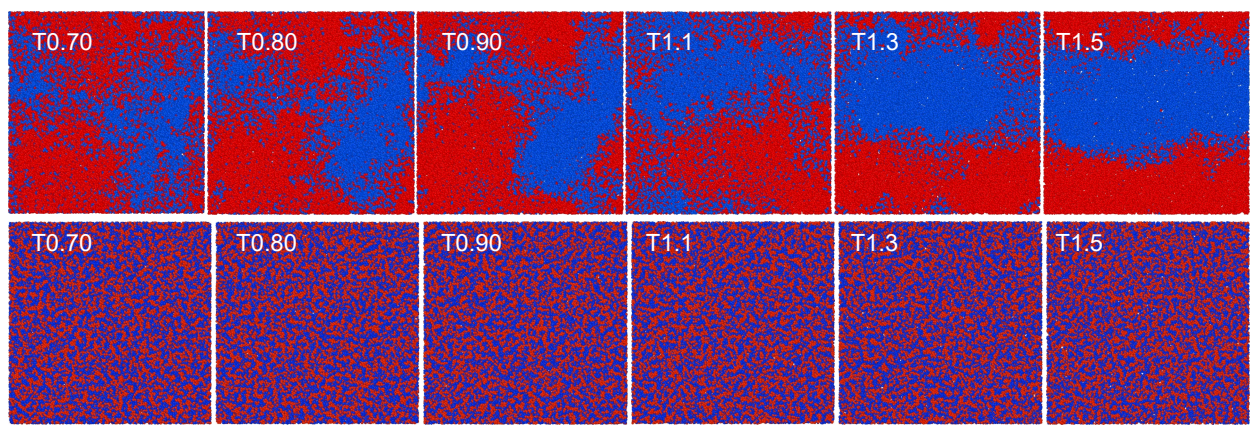

*Figure 5. Simulation snapshots of a hydrogen-bonding homopolymer blend (top row) and an alternating copolymer comprised of the same hydrogen bonding repeat units that comprised those homopolymers (bottom row). Snapshots are in order of increasing Lennard Jones temperature (temperatures shown in panels) from left to right. The homopolymer system's behavior is consistent with an LCST phase transition, which is suppressed in the alternating copolymer.*

As seen in Figure 4, in the case of unfavorable exchange energies, the fraction $f_{AB}$ considerably changes as the sequence approaches the alternating limit. In large blocks, $f_{AB}$ is small indicating few cross-species contacts; as the block length is lowered towards alternating, nonbonded opposite type interactions increase until the fraction, $f_{AB}$, is nearly half of the total nonbonded adjacencies. These findings are consistent with the prior work of Drayer et al[61]: in unfavorable exchange energy systems, increasing content of A-B bonds forces opposite-type monomers to mix despite their energetic penalty.

The impact of sequence on cross-adjacencies is shown by Figure 4 to be radically different in the case of favorable vdW exchange energies. Here, $f_{AB}$ is minimally perturbed regardless of block length. The largest deviation is seen in the alternating limit where $f_{AB}$ is reduced relative to the long-block limit. This seems surprising – why should increasing covalent interconnectivity between A and B repeat units *reduce* the number of nonbonded A-B contacts?

The reason why this occurs can be understood as follows. In a system for which the homopolymers (or long blocks) are immiscible, A-B bonds force A and B domains to come into close contact, thereby increasing the number of *nonbonded* cross interactions. In contrast, for a system with favorable A-B vdW interactions, A and B segments are already well mixed even in the long-block or homopolymer limits. Increasing the number of A-B covalent bonds then has the counterintuitive effect of merely replacing A-B nonbonded interactions with A-B bonded interactions. It follows that, of nonbonded adjacencies, a higher number are then AA or BB rather than AB – more A-B interactions are now bonded. Since the cohesive energy is determined by nonbonded interactions, this slightly lowers the cohesive energy and thus slightly lowers $T_g$ relative to the mixed homopolymers. This explains the modest reduction in $T_g$ relative to the homopolymer-blend limit shown in Figure 2 (but still appreciable enhancement relative to the pure polymer), and it explains the correspondingly weak reduction in $f_{AB}$ shown in Figure 4.

Notably, the $T_g$ sequence dependence of the simulated vdW system with favorable cross-interactions is qualitatively inverted from that observed experimentally in poly(vinylidene chloride-co-methyl acrylate).[51,52,85] In that system, $T_g$ is enhanced, but this enhancement is maximized in the alternating limit rather than the blockier limit. This poses an apparent dilemma. In light of the discussion above, the presence of a $T_g$ enhancement suggests preferentially favorable cross-interactions (i.e. favorable exchange energy) between monomers. However, the enhancement of this effect in the alternating limit suggests that A-B bonding enhances mixing, unlike in our simulations with favorable interactions. This implies that the experimental system exhibits a tendency towards thermodynamic demixing of the underlying homopolymers (at sufficiently high molecular weight) despite the presence of a favorable exchange energy (implying a negative chi). Under what circumstances can this occur?

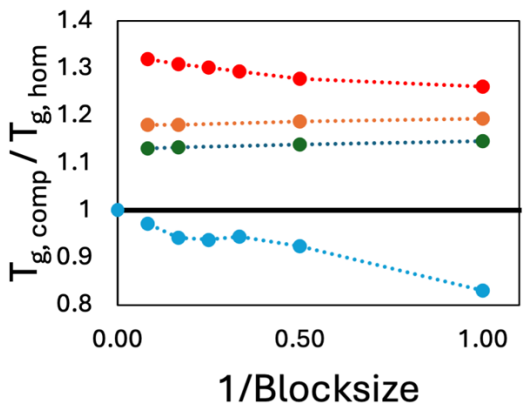

*Figure 6. Copolymer $T_g$ normalized by the homopolymer $T_g$, plotted vs inverse blocksize, for four systems: a system with unfavorable van der Waals interactions between the two monomer types, a system with favorable vdW interactions between the two monomer types, and two systems with model hydrogen bonds, of two different strengths, between the monomer types.*

## $T_g$ sequence effects in model hydrogen-bonding systems exhibiting LCST behavior

It follows from these observations that the behavior experimentally observed in poly(vinylidene chloride-co-methyl acrylate) likely occurs in the presence of LCST liquid-liquid phase equilibria behavior. LCST systems phase separate due to an unfavorable contribution to the entropy of mixing, despite a favorable energy of mixing – the exact opposite of the more intuitive UCST case.[75] This behavior commonly occurs in the presence of strong directional interactions, such as hydrogen bonds, between the species. Formation of these bonds on mixing is energetically favorable but comes with an entropic cost in terms of orientational degrees of freedom of the bonding groups.

To test this hypothesis, we implement a modified version of a bead-spring hydrogen bonding model that has been proposed in the literature[86] (Figure 1). As shown in Figure 5, the model we employ exhibits an LCST, as visually identified by progressive mixing of a homopolymer blend on *cooling*. This is the defining signature of an LCST. Crucially, the underlying homopolymer blend system remains at least partially phase separated over much of the glass formation range for these systems. This opens the possibility of realizing a system such as that hypothesized above, in which cross-interactions are favorable and yet this system tends towards demixing in the long-block limit in the glass formation range. Notably, integrating the two monomers into an alternating copolymer indeed suppresses this transition and yields a well-mixed system of A and B monomers, as shown in the lower row of Figure 5. This confirms that the system exhibits the phase behavior that we hypothesize leads to experimental sequence effects in $T_g$-enhancing copolymers: alternation enhances mixing despite the presence of strong favorable cross-interactions.

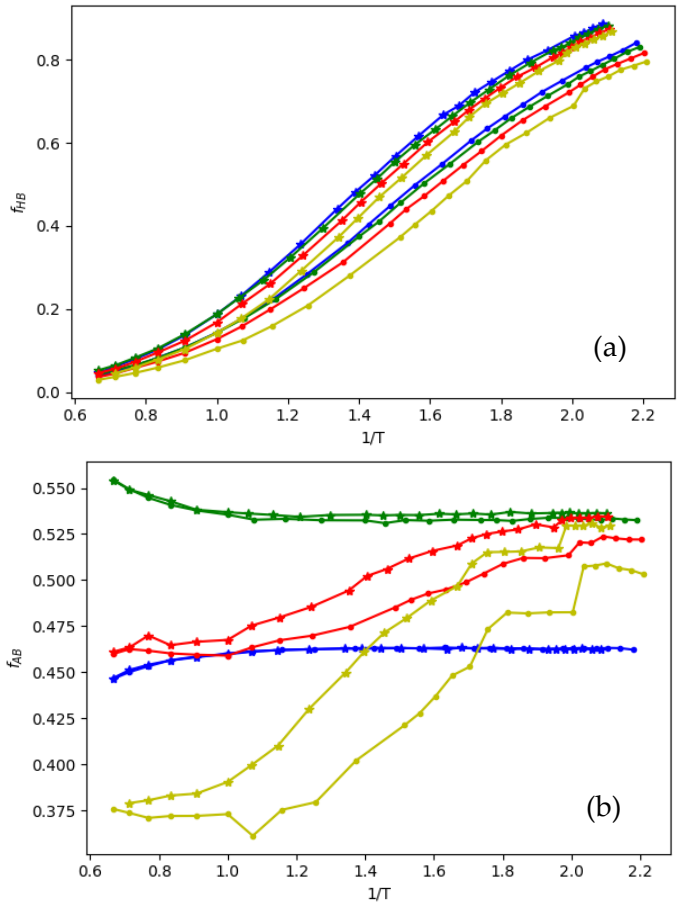

*Figure 7. Top: The fraction of possible hydrogen bonds formed in hydrogen bonding systems as a function of inverse temperature. Bottom: $f_{AB}$ for backbackbone interactions vs inverse temperature in hydrogen bonding systems. Datasets denoted by star symbols correspond to those with $\varepsilon_{AD} = 5.5$ and those denoted by circle symbols to those with $\varepsilon_{AD} = 5.0$. Datasets correspond to chains with repeating block size 12 (yellow), 6 (red), 2 (green) and 1 (blue) (bottom to top in upper panel.*

As shown in Figure 6, the presence of hydrogen bonds indeed dramatically alters the sequence dependence of $T_g$ in the copolymer. Systems with strong favorable cross hydrogen bonds between the monomers exhibit an enhancement in $T_g$ on mixing relative to Fox. Unsurprisingly, stronger cross hydrogen bonds lead to stronger $T_g$ enhancement, as stronger bonds lead to larger activation barriers for segmental relaxation. Unlike systems in which preferentially favorable cross-interactions have a vdW (dispersion force) nature, this enhancement grows upon approach to the alternating sequence limit.

This occurs for the reason hypothesized above: the hydrogen bonding systems exhibit entropically driven LCST phase separation due to the entropic cost of hydrogen bonding. As shown in Figure 7a, alternation thus increases the number of hydrogen bonds relative to the long-block limit. As a consequence of this increasing cohesive interaction strength, dynamics slow and $T_g$ then increases upon approach to alternation.

Notably, in these systems both the forced-mixing and the bond-replacement scenarios appear to be present. As shown in Figure 7b, shortening block sizes initially increases $f_{AB}$, indicating that more alternation induces a forced-mixing effect. However, $f_{AB}$ then decreases when reducing block lengths from alternating dimers to alternating monomers. This indicates that the alternating dimer case has approximately reached the entirely well mixed limit. Further increases in A-B bonds then begin to replace non-bonded A-B adjacencies with bonded A-B adjacencies. In this case, however, this does not produce a signature in $T_g$ vs block length (i.e. there is no nonmonotonicity wherein $T_g$ drops for the shortest blocks in Figure 6 for hydrogen-bonding systems), because hydrogen bonded interactions rather than simple van der Waals adjacencies are the most important contributor to activation energies for relaxation. Because formation of a hydrogen bond for a given segment requires only a *single* opposite type bead in the nearest neighbor shell, the differences in adjacency fractions shown in Figure 7b have little effect on hydrogen bonding densities and thus on relaxation times and $T_g$.

## Discussion and Conclusions

These results and analyses suggest that an interplay between liquid-liquid phase equilibria, energetic interactions, and sequence can govern deviations from the Fox rule in copolymers. In Figure 8, we schematically depict the dependence of $T_g$ sequence effects on liquid-liquid phase behavior that is suggested by our results. In the case of UCST systems, when $T_g$ lies near or within the underlying UCST of the two corresponding homopolymers, favorable cross-interactions suppress $T_g$ in the copolymer, and further suppression is found as sequence alternation induces more forced segmental mixing. When $T_g$ lies within or near an LCST phase boundary of the underlying homopolymers, this situation is precisely inverted: favorable cross-interaction energetics enhance $T_g$, with this enhancement growing in the alternating limit – again due to increased forced segmental mixing. Finally, we interpret the case of favorable vdW cross-interactions as corresponding to an LCST-like system, but wherein $T_g$ is far below any LCST temperature. In this case, copolymerization or mixing raises $T_g$, but this enhancement becomes relatively *muted* in the alternating limit, because A-B bonds reduce nonbonded A-B adjacencies in a system that begins well-mixed, thus reducing cohesive energy.

Practically, these findings provide guidance for the use of sequence control to modulate $T_g$ in copolymers. Copolymers comprised of UCST-prone copolymers (typically those with only van der Waals cross interactions) should be amenable to $T_g$ suppression through alternating copolymerization, or relative enhancement through blockier copolymerization. The inverse will be true for LCST-prone copolymers, which tend towards those with strong directional interactions such as hydrogen bonds between species. This explains why experimental copolymers such as poly(vinylidene chloride-co-methyl acrylate), which involve strong polar interactions between species, tend to exhibit $T_g$ enhancement that is augmented with increasing alternation.[37,51,52] Indeed, our recent atomistic simulations probing blends of PMMA with high-substitution chloroparaffins – a nearly chemically homologous system – identified the specific Cl-C partial charge interactions that are responsible for strong cohesive interactions in this system.[87]

This also explains why a copolymer of styrene and oligo-oxyethylene methacrylate was found to exhibit $T_g$ suppression and enhanced dynamics in a taper as opposed to a highly blocky configuration[88]. Oligo-oxyethylene methacrylate and styrene cross-interact primarily via weak van der Waals interactions, while their self-interactions involve polar interactions and pi-stacking interactions, respectively. This is thus a system where one would expect an effective Berthelot prefactor considerably less than one. This type of system tends to exhibit UCST behavior, and thus would be expected to exhibit $T_g$ suppression on mixing with further $T_g$ suppression as alternation increases. Since tapered copolymers involve far more A-B alternating-like bonds than do blocky copolymers, it follows that the tapers should exhibit a lower $T_g$ and higher mobility than blocks.

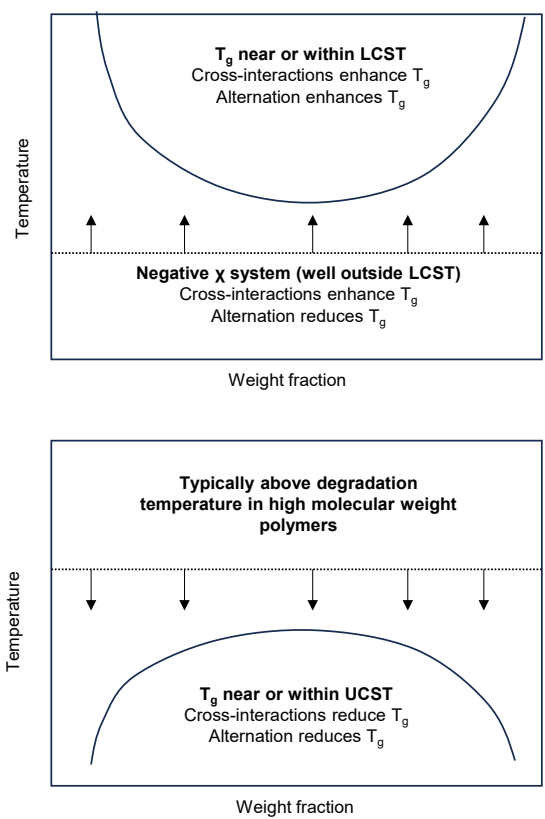

*Figure 8. Schematic depicting $T_g$ interaction and sequence effects for LCST polymers (top) and UCST polymers (bottom) based on the results of this work. The phase boundaries correspond to the qualitative tendency of the constituent monomers and their oligomers and polymers towards liquid-liquid phase separation in the homopolymer-blend limit.*

In summary, these findings suggest a clear scenario wherein sequence control can be leveraged to enhance or reduce the $T_g$ of high-performance copolymers. As shown by the schematic in Figure 8, our findings suggest that interaction- and sequence-mediated $T_g$ deviations from simple copolymer $T_g$ mixing rules can be understood and qualitatively predicted based on the location of the homopolymer $T_g$ within the liquid-liquid phase diagram of the homopolymer (or oligomer) blend. This behavior fundamentally emerges from a forced mixing phenomenon wherein monomer alternation induces close contact between monomer types that would otherwise tend towards phase separation.

### Acknowledgements

The authors acknowledge funding support from the Air Force Office of Scientific Research under grant number FA9550-22-1-0427.

### Author Contributions

MRB performed and analyzed all simulations under DSS' supervision. All authors contributed to writing the manuscript.

### Competing Interests

The authors declare no competing interests.

### Additional Information

Supplementary relaxation time data are available free of charge in the Supplementary Data file